\documentclass{iopart}

\usepackage{epsfig}
\begin{document}
\def\b{\bar}
\def\d{\partial}
\def\D{\Delta}
\def\cD{{\cal D}}
\def\cK{{\cal K}}
\def\f{\varphi}
\def\g{\gamma}
\def\G{\Gamma}
\def\l{\lambda}
\def\L{\Lambda}
\def\M{{\Cal M}}
\def\m{\mu}
\def\n{\nu}
\def\p{\psi}
\def\q{\b q}
\def\r{\rho}
\def\t{\tau}
\def\x{\phi}
\def\X{\~\xi}
\def\~{\widetilde}
\def\h{\eta}
\def\bZ{\bar Z}
\def\cY{\bar Y}
\def\bY3{\bar Y_{,3}}
\def\Y3{Y_{,3}}
\def\z{\zeta}
\def\Z{{\b\zeta}}
\def\Y{{\bar Y}}
\def\cZ{{\bar Z}}
\def\`{\dot}
\def\be{\begin{equation}}
\def\ee{\end{equation}}
\def\bea{\begin{eqnarray}}
\def\eea{\end{eqnarray}}
\def\half{\frac{1}{2}}
\def\fn{\footnote}
\def\bh{black hole \ }
\def\cL{{\cal L}}
\def\cH{{\cal H}}
\def\cF{{\cal F}}
\def\cP{{\cal P}}
\def\cM{{\cal M}}
\def\olam{\stackrel{\circ}{\lambda}}
\def\oX{\stackrel{\circ}{X}}
\def\const{{\rm const.\ }}
\def\ik{ik}
\def\mn{{\mu\nu}}
\def\a{\alpha}

\title[Complex Kerr Geometry, Twistors and the Dirac Electron]{
 Complex Kerr Geometry, Twistors and the Dirac Electron}

\author{A Burinskii}

\address{Gravity Research Group NSI Russian Academy of Sciences.
B. Tulskaya 52, Moscow 115191, Russia} \ead{bur@ibrae.ac.ru}
\begin{abstract}
The Kerr-Newman spinning particle displays some remarkable
relations to the Dirac electron and has a reach spinor structure
which is based on a twistorial description of the Kerr congruence
determined by the Kerr theorem. We consider the relation between
this spinor-twistorial structure and spinors of the Dirac
equation, and show that the Dirac equation may naturally be
incorporated into Kerr-Schild formalism as a master equation
controlling the twistorial structure of Kerr geometry. As a
result, the Dirac electron acquires an extended space-time
structure having clear coordinate description with natural
incorporation of a gravitational field. The relation between the
Dirac wave function and Kerr geometry is realized via a chain of
links: {\it Dirac wave function $ \Rightarrow $ Complex
Kerr-Newman Source $ \Rightarrow $ Kerr Theorem $ \Rightarrow $
Real Kerr geometry.} As a result, the wave function acquires the
role of an ``order parameter'' which controls spin, dynamics, and
twistorial polarization of Kerr-Newman space-time.
\end{abstract}


\section{Introduction}
The fact that Kerr-Newman solution has gyromagnetic ratio $g=2$ as
that of  the Dirac electron \cite{Car} created the treatment of
this solution as a classical model of an extended electron in
general relativity
\cite{Car,Isr,DKS,Bur0,IvBur,Bur1,IvBur1,Lop,BurStr,BurSup,BurOri,ArcPer,BurTwi,Pek}.
If this coincidence is not occasional, one has to answer a
fundamental question: what is the relation of the Dirac equation
to the structure of the Kerr-Newman solution? Contrary to the
Dirac electron, the Kerr-Newman spinning particle has clear
space-time structure which is concordant with gravitational field.

One can argue that the gravitational field of an electron is
negligibly weak and can be ignored. However, one cannot ignore the
extremely large spin/mass ratio (about $10^{44}$ in the units
$\hbar=c=G=1 ,$) which shows that the correct estimations of the
gravitational effects have to be based on the Kerr-Newman
solution. Results of corresponding analysis are rather unexpected
\cite{BurReg,Ros} and differ drastically from the estimations
performed on the base of spherically symmetric solutions.
 Although the local averaged gravity is  very weak,
the extremely high spin leads to the very strong polarization of
space-time and to the corresponding very strong deformation of
electromagnetic (em-) field which has to be aligned with the Kerr
congruence. Since the em-field of electron cannot be considered as
small, the resulting influence turns out to be very essential. In
particular, the em-field turns out to be singular at the Kerr
singular ring which has the Compton size.
 Moreover, the Kerr-Newman space-time with
parameters of an electron is topologically not equivalent to the
flat Minkowski space-time, acquiring two folds with a branch line
along the Kerr ring. It shows that the Kerr geometry gives some
new background for the treatment of this problem. In fact, the
Kerr solution gives us some complementary coordinate information
which has a natural relation to gravity and displays independently
the special role of the Compton region.

The aim of this paper is to set an exact correspondence between
the operators of polarization and momentum of an electron in the
Dirac theory and similar relations for the momentum and spin of
the Kerr spinning particle:

\be Dirac \ equation  \ \Rightarrow \ wave \ function  \
\Rightarrow \  Kerr \ Geometry \label{Ch0}. \ee

As a result, we obtain a model, in which electron has the extended
space-time structure of Kerr-Newman geometry and {\it the Dirac
equation is considered as a master equation controlling dynamics
and polarization of this structure.}\fn{In fact we arrive at some
stochastic version of one-particle quantum theory with hidden
structure, similar to theories with hidden parameters.}

Our treatment is based on the initiated by Newman \cite{New}
complex representation of the Kerr geometry, in which a
`point-like' source of the Kerr-Newman solution is placed in a
complex region and propagates along a complex world-line
$X^\m(\t)$ in a complexified Minkowski space-time $CM^4.$

It was shown \cite{BurKer,BurNst} that a natural and rigorous
treatment of this construction may only be achieved in the
Kerr-Schild formalism \cite{DKS} which is based on the metric
decomposition $g_\mn =\eta_\mn - 2 H k_\m k_\n$ containing
auxiliary Minkowski space-time $M^4$ with metric $\eta _\mn .$
This auxiliary $M^4$ is complexified to $CM^4$ and can be used as
a natural space-time for the complex Kerr source as well as for
the Kerr null vector field $k^\m(x), \ ( x=x^\m \in M^4 ),$
forming the Kerr congruence via  Kerr theorem
\cite{BurKer,BurNst,Pen,KraSte}.

The Kerr theorem determines the Kerr congruence in $M^4$ from a
holomorphic generating function $F(Z)$ in terms of projective
twistor coordinates $Z^\alpha$. The relation of the Kerr geometry
to twistors is not seen in Boyer-Lindquist coordinates, but it
turns out to be profound in the Kerr-Schild formalism. Although
the terms `twistor' and `Kerr theorem' were absent in the seminal
paper \cite{DKS}, they were practically used there for derivation
of the Kerr-Schild class of solutions via the chain of relations

\be F(Z) \Rightarrow Y(x) \Rightarrow k^\m \Rightarrow g_\mn
\label{Ch1} ,\ee where the twistor coordinates \be Z^\alpha =(Y,
\quad \z - Y v, \quad u + Y \Z \ ) \label{PTw} \ee are  defined
via the null Cartesian coordinates \be 2^{1\over2}\z = x + i y
,\quad 2^{1\over2} \Z = x - i y , \quad 2^{1\over2}u = z - t
,\quad 2^{1\over2}v = z + t . \label{null}\ee The variable $Y$
plays a special role, being the projective spinor coordinate $ Y=
\phi^2/\phi^1 $ and, simultaneously, the projective angular
coordinate \be Y=e^{i\phi}\tan \frac \theta 2 \label{Y0} .\ee
Therefore, the output of the Kerr theorem, function $Y(x),$
determines the field of null directions $k^\m(x^\m)$ in $M^4 .$
This field forms a vortex of twisting null congruence, each
geodesic line of which represents the twistor $Z^\alpha =const$.

To match the Dirac solutions with the Kerr congruence we select
two special twistor lines going via the center of the solution,
$(t,x,y,z)=(0,0,0,0),$ and corresponding to $Y=0$ and $Y=\infty.$
 For the Kerr solution in a standard position, these lines form
 two semi-infinite axial beams directed along the positive ($\theta=0$)
 and negative ($\theta=\pi $) z-axis. They are determined by two
two-component spinors which we set corresponding to the
four-component Dirac spinor of the wave function of the Dirac
equation in a Weyl basis. In this case the spin-polarization and
momentum of the Dirac electron matches with the spin and momentum
of the Kerr spinning particle \cite{BurTwi,BurAxi}

Such a relation is simple for the standard orientation of the
Kerr-Newman solution and the standard treatment of the Dirac
equation for a free electron. However, in the more general cases
(a moving electron, or electron in an external electromagnetic
field) the null vectors formed by the Dirac bispinor components
turn out to be independent, and the corresponding selected null
beams take independent orientation. It leads  to the deformation
of the Kerr twistorial structure which is determined by some
generating function of the Kerr theorem. Thus, to set the
correspondence (\ref{Ch0}) in a general case, one has to use the
Kerr theorem involving the complex-world-line (CWL) representation
of Kerr geometry.
  The corresponding chain of relations takes the form $$ Dirac
\ equation  \ \Rightarrow \ wave \ function  \ \Rightarrow \  CWL
\Rightarrow \ F_q \ \Rightarrow \ Y(x) \Rightarrow \  Kerr \
Geometry $$ which is valid for a weak and slowly varying
electromagnetic field, which is the case compatible with the
validity of one-particle Dirac theory. The relation between CWL
and parameters $q$ of the Kerr
 generating functions $F_q$ was investigated in
 \cite{BurNst,BurMag}, and in this paper we consider
 the missing link $ \quad Dirac
\ equation  \ \Rightarrow \ wave \ function  \ \Rightarrow \  CWL
 \ $ which allows one to consider Dirac equation as a {\it master
equation,} controlling the polarization and dynamics of the Kerr
geometry corresponding to the wave function of the considered
electron.

One more aspect of our treatment concerns the wave properties of
electron. It was obtained long ago that stationary Kerr-Newman
solution may only be considered as a first approximation, and some
extra electromagnetic and spinor wave excitations on the Kerr
background are necessary to generate the wave properties of the
Dirac electron. Because of that, from the beginning this model was
considered as a model of `microgeon with spin' \cite{Bur0}, in
which the Kerr-Newman solution represents only solution for some
averaged fields on the Kerr background. It was observed
\cite{BurAxi,BurSup} that the treatment of the electromagnetic or
spinor wave excitations on the Kerr background leads to the
inevitable appearance of extra axial singular lines resembling the
singular strings of the Dirac monopole. Moreover, the wave
excitations of the Kerr circular singularity induce de Broglie
periodicity on the axial singular lines \cite{BurTwi,BurAxi}. It
stimulated investigation of the wave analogs of the Kerr-Newman
solutions, which is related with generalization of the known
Kerr-Schild class of solutions by treatment of some extra function
$\gamma$ which was set to zero in the general Kerr-Schild
formalism. This is a very hard unsolved problem, and  in this
paper we concentrate attention on special exact singular solutions
`chirons' which acquire a wave generalization, being
asymptotically exact  for the weak and slowly varying
electromagnetic excitations \cite{BEHM1}.

We keep mainly the Kerr-Schild notations \cite{DKS} for Kerr
geometry and spinor notations of the book \cite{BLP}.
 The following two sections represent a brief description of the
structure of Kerr solution following the papers
\cite{BurOri,BurTwi,BurNst}.

\section{Real structure of the Kerr geometry}
The angular momentum of an electron $J=\hbar /2$  is extremely
high with respect to the mass, and the black hole {\it horizons
disappear,} opening the naked Kerr singular ring. This ring is a
branch line of the space which acquires two-fold topology. It was
suggested \cite{Bur0} that the Kerr singular ring represents a
string which may have some excitations generating the spin and
mass of the extended particle-like object -- `microgeon'.

The skeleton of the Kerr geometry is formed by {\it the Kerr
principal null congruence} which represents a twisted family of
the lightlike rays -- twistors. The null vector field $k^\m (x)$,
which is tangent to these rays, determines the Kerr-Schild form of
metric \be g^\mn =\eta^\mn + 2H k^\m k^\n \label{KS}, \ee where
$\eta^\mn$ is the auxiliary Minkowski metric with coordinates
$x^\m =(t,x,y,z).$ The vector potential of the Kerr-Newman
solution is aligned with this congruence

\be A_\m = e r (r^2+ a^2 \cos ^2 \theta)^{-1} k_\m ,
\label{Aem},\ee and the Kerr singular ring represents its caustic.

{\bf The Kerr theorem} \cite{Pen,KraSte,BurKer,BurNst,Multiks}
claims that {\it any holomorphic surface in the projective twistor
space $CP^3$} with coordinates
 \be Z^\alpha =(Y, \ \l^1 , \ \l^2 ), \quad
 \l^1 = \z - Y v , \quad  \l^2 =u + Y \Z \      \label{Tw} \ee
 determines the geodesic and shear-free null congruence in
$M^4.$ Such congruences lead to solutions of the Einstein-Maxwell
field equations with metric (\ref{KS}) and an em-field in the form
(\ref{Aem}). The congruence of the Kerr solution is built of the
straight null generators, twistors, which are (twisting) null
geodesic lines (possible trajectories of photons).
\begin{figure}[ht]
\centerline{\epsfig{figure=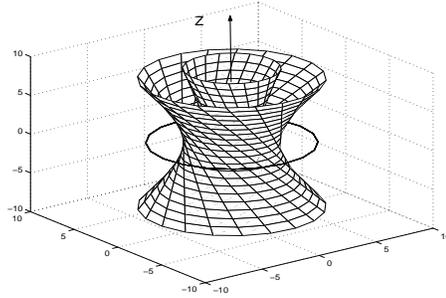,height=4cm,width=6cm}}
\caption{The Kerr singular ring and  congruence. }
\end{figure}
Therefore, for any holomorphic function $F(Z^\alpha) ,$ solution
$Y(x^\m)$ of the equation $F(Y,\l_1,\l_2)=0 $ determines
congruence of null lines in $M^4$ by the form \be e^3= du + \bar Y
d \zeta + Y d \bar\zeta - Y \bar Y dv \label{e3} , \ee  and the
null vector field tangent to congruence is $k_\m dx^\m=
 P^{-1}e^3 .$  \fn{Here $k^\m$ and $Y$ are functions of $x^\m =(t,x,y,z) \in M^4 ,$
 and $P=P(Y,\Y)$ is a normalizing
 factor related with the boost of  Kerr source.}
 Function $Y$ is related to projective
 spinor, $Y=\phi_2/\phi_1,$ and null vector field $k^\m$ may be represented
 in spinor form $k_\m = \bar  \phi _{\dot\alpha} \bar\sigma
_\m^{\dot\alpha\alpha}  \phi _\alpha .$

\section{Complex representation of the Kerr geometry }
{\bf The complex source of Kerr geometry} is obtained as a result
of complex shift of the `point-like' source of the Schwarzschild
solution in the Kerr-Schild form. There are also the Coulomb and
Newton analogs of the Kerr solution.

Applying the complex shift $(x,y,z) \to (x,y,z+ia)$ to the
singular source $(x_0,y_0,z_0)=(0,0,0)$ of the Coulomb solution
$q/r$, Appel in 1887(!) obtained the solution $ \phi(x,y,z)= \Re e
\ q/\tilde r, $
 where $\tilde r =\sqrt{x^2+y^2+(z-ia)^2}$ turns out to be
complex. On the real slice $(x,y,z)$, this solution acquires a
singular ring corresponding to $\tilde r=0.$ It has radius $a$ and
lies in the plane $z=0.$  The solution is conveniently described
in the oblate spheroidal coordinate system $r, \ \theta,$ where
\be \tilde r =r+ia\cos\theta .\label{tr} \ee One can see that the
space is twofold having the ring-like singularity at
$r=\cos\theta=0$ as the branch line. Therefore, for the each real
point $(t,x,y,z) \in {\bf M^4}$ we have two points, one of them is
lying on the positive sheet, corresponding to $r>0$,  and another
one lies on the negative sheet, where $r<0$.

It was obtained that Appel potential corresponds  exactly to
electromagnetic field of the Kerr-Newman solution written in the
Kerr-Schild form, \cite{Bur0}. The vector of complex shift $\vec
a=(a_x,a_y,a_z)$ corresponds to the angular momentum of the Kerr
solution.

Newman and Lind \cite{New} suggested a description of the
Kerr-Newman geometry  in the form of a retarded-time construction,
where it is generated by a complex point-like source, propagating
along a {\it complex world line} $X^\m(\t)$ in a complexified
Minkowski space-time $\mathbf{CM}^4$. The rigorous substantiation
of this representation is possible only in the Kerr-Schild
approach \cite{DKS} based on the Kerr theorem and the Kerr-Schild
form of metric (\ref{KS}) which are related to the auxiliary
$\mathbf{CM}^4$ \cite{BurNst,BurKer,BurMag}.

In the rest frame of the Kerr particle, one can form two null
4-vectors $k_L=(1,0,0, 1)$ and $k_R=(1,0,0, -1),$
 and  represent the 3-vector of complex shift $ i\vec a=i \Im m X^\m$ as the
  difference $ i\vec a =\frac{ia}{2} \{ k_L -
k_R \}.$  The straight complex world line corresponding to a free
particle may be decomposed to the form

\be X^\m(\t) = X^\m(0) + \t u^\m + \frac{ia}{2} \{ k_L - k_R \}
,\label{cwl}\ee

where the time-like 4-vector of velocity $u^\m=(1,0,0,0)$
 can also be represented via vectors $k_L$ and $k_R ,$

\be u^\m =\d _t \Re e X^\m(\t)=\frac 12 \{ k_L+k_R \}. \label{umu}
\ee
 One can form two complex world lines related to the complex Kerr
source, \be X_{+}^\m(t+ia) = \Re e X^\m(\t) + ia k^\m_L , \quad
X_{-}^\m(t-ia) = \Re e X^\m(\t) - i a k^\m_R , \label{X+} \ee
which allows us to match the Kerr geometry to the solutions of the
Dirac equation.

\subsection{Complex Kerr string}

The complex world line $X^\m (\t)$ is parametrized by the complex
time parameter $\t=t+i\sigma$ and represents a  world sheet.
Therefore, $X^\m (t,\sigma)$ is a very specific string extended
along the imaginary time parameter $\sigma$. The Kerr-Newman null
congruence and corresponding gravitational and electromagnetic
fields are obtained from this string-like source by a
retarded-time construction which is based on the complex null
cones, emanated from the worldsheet of this complex string
\cite{BurStr,BurTwi}. In particular, the real twistors of the Kerr
congruence represent a real slice of the null generators of these
null cones \cite{BurStr,BurNst}.  The complex retarded time
equation  $t - \t = \tilde r $ takes the form

\be \t = t - r  + ia \cos \theta . \label{tau} \ee The real
sections of the complex cones correspond to the real coordinates
$t,r, \theta .$ It yields the relation \be \sigma =a\cos \theta
\label{cos} \ee between the points of worldsheet and angular
directions of twistor lines. Since $|\cos \theta |\le 1 ,$ we
conclude that the string is open and has the end points,
corresponding to $\cos\theta= \pm 1$ and to two complex world
lines $X_{+}^\m=X^\m(t+ia)$ and $X_{-}^\m=X^\m(t-ia). $ By
analogue with the real strings, where the end points are attached
to quarks, one can add the Chan-Paton factors to the end points
$X_{\pm}^\m$ of the complex Kerr string and identify them as
quarks \cite{BurOri,BurTwi}.

The complex cones positioned at these end points have the {\it
real slice in the form of two real twistors} corresponding  with
the above discussed null directions $k_L^\m$ and $k_R^\m $ which
determine momentum and spin-polarization of the Kerr solution.
These twistors have the limiting values of angular direction
$\cos\theta= \pm 1 ,$ and form two half-strings of opposite
chirality aligned with the axis of symmetry z.
\subsection{Chirons and excitations of the Kerr singular ring}
The twistor coordinate $Y$ is also the projective angular
coordinate \be Y=e^{i\phi}\tan \frac \theta 2 \label{Y}\ee
covering the celestial sphere $Y\in CP^1 =S^2.$
\begin{figure}[ht]
\begin{center}
\psfig{file=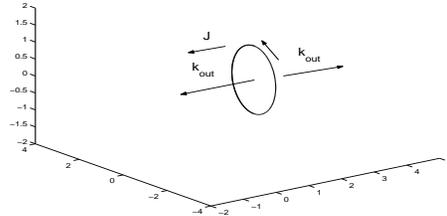,height=3.5cm,width=6cm}
\end{center}
\caption{The Kerr singular ring and two special twistors.}
\end{figure}
 Electromagnetic field of the exact stationary Kerr-Schild solutions,
 \cite{DKS},
 is determined by vector potential which may be represented in the
 form
 \be A^\m =  \Re e \ {\cal A}(Y)(r+ia\cos \theta)^{-1} k^\m ,\ee
 where ${\cal A}(Y)$ is an arbitrary analytical function of $Y $.
 In general, ${\cal A}(Y)$ may
 contain the poles in different angular directions $Y$ of the celestial sphere
 $S^2$ , which causes the appearance of  semi-infinite singular rays -- axial
 strings \cite{BurAxi}.  The elementary solutions are ${\cal A}(Y) = e Y^{-n} .$
The simplest case $\psi= e = const.$ gives
 the Kerr-Newman solution.
 The case $n=1$ leads to
 a singular line along the positive semi-axis z. Due to the factor $e^{i\phi}$ in
$Y ,$ the em-field has winding number n=1 around this axial
singularity. Since
 there is also pole at singular ring,  $\sim (r +ia \cos\theta)^{-1},$
 the em-field has a winding of phase along the Kerr ring.
 The solution with $n=-1$ has opposite chirality and singular line along the
 negative semi-axis z. The elementary exact solutions (`chirons') have
 also the wave generalizations
 ${\cal A} =eY^{-n}e^{i\omega\t}$ acquiring the extra dependence from the
 complex retarded time $\t ,$ \cite{BurAxi}. The wave chirons are
 asymptotically exact in the low-frequency limit \cite{BEHM1}
and describe the waves propagating along the Kerr singular ring,
as it was assumed in the old `microgeon' model \cite{Bur0}. Such
waves may also be considered as em-excitations of the Kerr closed
string \cite{BurOri}. The two axial half-strings are not
independent: the boundary conditions of the complex Kerr string
demands its orientifolding  by identification of the initiate
worldsheet and the worldsheet with reverse parametrization
\cite{BurStr,BurTwi}. Orientifolding is accompanied by the revers
of space and antipodal map $\Y \to -1/Y ,$ which displays
antipodal relation between the singular half-strings and also
between the corresponding chirons.   Notice that by a lorentz
boost the axial half-strings acquire modulation by de Broglie
periodicity \cite{BurAxi,BurTwi}.

\section{Dirac equation in the Weyl basis}
 In the Weyl basis  the Dirac spinor has the form
 $\Psi =
\left(\begin{array}{c}
 \phi _\alpha \\
\chi ^{\dot \alpha}
\end{array} \right),$
and the Dirac equation splits into \be\sigma ^\m _{\alpha \dot
\alpha} (i \d_\m  +e A_\m)
 \chi ^{\dot \alpha}=  m \phi _\alpha , \quad
 \bar\sigma ^{\m \dot\alpha \alpha} (i \d_\m  +e A_\m)
 \phi _{\alpha} =  m \chi ^{\dot \alpha}.\label{Wspl} \ee
 The conjugate spinor has the form
 \be \bar\Psi =(\chi ^+,\phi ^+ )=(\bar \chi^\alpha, \bar
 \phi_{\dot\alpha}). \ee

The Dirac current \be J_\m = e (\bar \Psi \gamma _\m \Psi) = e
(\bar\chi  \sigma _\m  \chi + \bar\phi  \bar \sigma _\m  \phi ),
\ee  can be represented as a sum of two lightlike components of
opposite chirality \be J^\m_{L} = e \bar\chi \sigma^\m \chi \ ,
\qquad J^\m_{R} = e \bar\phi \bar\sigma^\m \phi. \ee

The corresponding null vectors

\be k^\m_{L} = \bar\chi \sigma^\m \chi \ , \quad k^\m_{R} =
\bar\phi \bar\sigma^\m \phi , \label{kLR} \ee determine the
considered above directions of the lightlike half-strings. The
momentum of the Dirac electron is $p^\m = \frac m 2 (k^\m_{L} +
k^\m_{R}),$ and the vector of polarization of an electron
\cite{AkhBer,BLP} in the state with a definite projection of spin
on the axis of polarization  is $n^\m = \frac 12 (k^\m_{L} -
k^\m_{R}).$ In particular, in the rest frame and the axial
z-symmetry, we have $k_{L}=(1,\vec k_{L})=(1,0,0,1)$ and
$k_{R}=(1,\vec k_{R})=(1,0,0,-1),$ which gives
 $p^\m = m(1,0,0,0),$ and $n^\m =  (0,0,0,1),$
which corresponds to the so-called transverse polarization of
electron \cite{AkhBer}, $\vec n \vec p =0 .$

By the Lorentz boost $\vec v ,$ the spinors $\chi$ and $\phi$
transform independently \cite{Fey} \be \chi '=\exp (-\sigma_v
\frac w 2)\chi, \quad \phi '=\exp (-\sigma_v \frac w 2)\phi ,\ee
where $\sigma_v= (\vec \sigma \cdot \vec v)/|v| $ and $\tanh w=
v/c.$ The Dirac spinors form a natural null tetrad. The null
vectors $ k^\m_{L} = \bar\chi \sigma^\m \chi $ and $ k^\m_{R} =
\bar\phi \bar\sigma^\m \phi $, may be completed to the null tetrad
by two null vectors $ m^\m = \phi \sigma^\m \chi \ ,$ and $\bar
m^\m = (\phi \sigma^\m \chi)^+ $ which are controlled by the phase
of wave function. Therefore, the de Broglie wave sets a
synchronization of the null tetrad in the surrounding space-time,
playing the role of an `order parameter'.

\section{Dirac equation as a master equation controlling
twistorial polarization}

Obtaining the relation between the Dirac wave function and Complex
World Line (CWL), we have to set other missing links in our long
chain of the relations discussed in the introduction. First, let
us recall the relation $F \Rightarrow Kerr \ Geometry .$ The used
in \cite{DKS} generating function $F(Y,\l^1,\l^2)$ leading to the
Kerr-Newman solution had the form

\be F \equiv a_0 +a_1 Y + a_2 Y^2 + (q Y + c) \l ^1 - (p Y + \q)
\l ^2, \label{FK} \ee where $a_0, a_1, a_2$  are complex constants
which determine spin orientation, the coefficients $ c, p, , q,
\q, $ determine the Killing vector (or the boost) of the solution
and the related function \be P= -pY\Y - \q \Y - q Y - c \ .
\label{Ppc} \ee

Since $ \l^1 = \z - Y v , \quad  \l^2 =u + Y \Z ,$ function $F$ is
quadratic in $Y$ and has the general form $ F = A  Y^2 + B Y + C,$
which allows to find two roots of the equation $F(Y) = 0 \ , $

\be Y^\pm = (- B + \D^\pm )/2A, \quad \D ^\pm = \pm (B^2 -
4AC)^{1/2} , \label{Ypm}\ee

and represent function $ F(Y) $  in the form

\be F=A(Y-Y^+)(Y-Y^-) . \label{FYpm}\ee

Following \cite{DKS,BurNst} we can determine  \be PZ^{-1} = - \d F
/\d Y= 2AY + B, \label{tr2} \ee which turns out to be a complex
radial distance $\tilde r = r+ia\cos\theta ,$

\be \tilde r ^{\pm}=-PZ^{-1} = 2AY^{\pm} + B = \D ^\pm .
\label{trpm} \ee

 The two solutions for $Y$ and $\tilde r$ reflect the known twofoldedness of the real Kerr
geometry and correspond to two different sheets of the real Kerr
space-time with different congruences.

In the case of arbitrary position, spin orientation and boost, the
generating function $F$ is controlled by the set of parameters $q
=(a_0, a_1, a_2, c, q, \q, p)$ and may be represented as $F_q= A_q
Y^2+B_q Y +C_q .$ The relation of the coefficients $A,B,C$ with
parameters of CWL (\ref{cwl}) was given in \cite{BurNst,BurMag}:
\bea
 A &=& (\Z  - \Z_0) \`v_0 - (v-v_0) \`\Z_0 ;\nonumber\\
 B &=& (u-u_0) \`v_0 + (\z - \z_0 )\`\Z_0
  - (\Z - \Z_0) \`\z_0 - (v - v_0) \`u_0 ;\nonumber\\
C &=& (\z - \z_0 ) \`u_0 - (u -u_0) \`\z_0, \eea

were the parameters of CWL are expressed in the null coordinates
(\ref{null}), in accordance with the correspondence $$ (u_0, v_0,
\z_0, \Z_0) \leftrightarrow X^\m(0) +\frac{ia}{2} \{ k_L - k_R \}
, \quad (\`u_0, \`v_0, \`\z_0, \`\Z_0) \leftrightarrow
\`X^\m(\t)=\frac 12 \{ k_L+k_R \}.$$

It restores the full chain of relations between the values of the
Dirac wave function and polarization of the Kerr geometry.

The obtained relationship $Dirac \ theory \Rightarrow Kerr \
geometry $ may be interpreted in the frame of some version of
one-particle quantum theory. For example, the plane Dirac wave
does not give information on the position of electron, but gives
exact information on its momentum and spin-polarization. The
center of corresponding Kerr-Newman geometry may be localized at
any point of space-time, but the considered model gives a definite
orientation of spin and corresponding deformation of the Kerr
congruence caused by momentum (Lorentz boost). If the wave
function is formed by a wave packet and localized in some
restricted region which is much greater than the Compton length,
we have a value of the normalized Dirac bispinor $\Psi =
\left(\begin{array}{c}
 \phi _\alpha (x)\\
\chi ^{\dot \alpha}(x)
\end{array} \right),$ at the point $ x\in M^4$ and
corresponding density of probability $w(x)=\bar\chi (x) \chi (x) +
\bar\phi(x) \phi (x)$ for position of the Kerr geometry, and  the
obtained relationships say that with the density $ w(x)$ electron
is positioned at this point and has at this point a definite
polarization of the Kerr-Newman geometry which is determined by
values $\chi(x)$ and $\phi (x)$. Therefore, we arrive at some
stochastic version of quantum theory containing hidden parameters,
more precisely -- a hidden space-time structure.

The relationship $Dirac \ theory \Rightarrow Kerr \ geometry $ is
one-sided, since it regards the {\it Dirac equation as a master
equation} and does not give anything new for the Dirac theory
itself besides of its interpretation. At the same time it gives
some new useful relations to Kerr-Newman solution, allowing us to
determine its behavior  in a weak and slowly varying external
electromagnetic field via the solutions of  Dirac equation.

On the other hand, the  considered twistorial structure of
electron is based on the local field theory and allows one to
conjecture that there is indeed a relation of this model to
multi-particle quantum field theory which gives a more detailed
description of electron. The author expects that at least some of
the mysteries and problems of the modern QED may be understood and
cured in this way. In particular, the em-field of the Kerr-Schild
solutions $F_{\mn}$ is to be aligned with the Kerr congruence,
obeying the constraint $F_{\mn}k^\m =0.$
 The twistorial structure of the Kerr-Schild solutions determines
polarization of the em field, providing a caustic on the Kerr
singular ring. Consequently, elementary electromagnetic excitation
aligned with the Kerr background shall lead to the appearance of
waves propagating along the Kerr ring and, simultaneously, to the
appearance of the induced singular axial pp-waves  \cite{BurAxi}.
All that has to be also valid for the vacuum fluctuations
\cite{BEHM1}, and the field of virtual photons is to be
concentrated near the Kerr singular ring, forming excitation of
this ring which may be considered as a closed string.\fn{It was
shown that the fields around the Kerr string are similar to the
fields around a heterotic string obtained by Sen as a solution to
low-energy string theory \cite{BurSen,BurOri}. However, it has the
peculiarity of `Alice' string, since it is a branch line of the
space onto `negative' and`positive' sheets, forming a gate to the
mirror `Alice' world.} Therefore, the model of electron based on
the fields aligned with the Kerr twistorial structure supports the
conjecture that the string-like source of the Dirac-Kerr electron,
having the Compton size, should be {\it experimentally
observable.}

It seems that there is a more simple way to apply Kerr geometry,
considering the Dirac equation on the Kerr background \cite{Pek}.
However, in spite of the separability of Dirac equation on the
Kerr background, the corresponding exact wave solutions are
unknown for the case of nonzero mass term. There are also some
theoretical arguments sowing that the exact massive solutions on
the Kerr background, aligned with the Kerr congruence, don't exist
at all, because of the twosheetedness of the Kerr space-time.

It should also be noted, that all the wave em-solutions aligned
with the Kerr background demonstrate the appearance of singular
beams. Such singular beams also appear inescapably in the spinor
wave solutions  \cite{BurSup,Pek}, which  disables the necessary
normalization of the wave functions. All that shows us the serious
problems with a straight approach and justifies the reason for the
treatment of the above combined Dirac-Kerr model. Although it is
apparently not a unique possible model. One of the other
prospective approaches we intend to consider is the treatment of a
local field theory based on the stringy character of the complex
Kerr source. In such a model the initially massless Dirac field is
related with the sources distributed over the worldsheet of the
complex Kerr string. In fact, the twistor lines adjoined to each
point of the considered above complex Kerr string may be carriers
of the massless Dirac field,  forming a sort of twistor-string
\cite{BurTwi,Wit}. The massive Dirac equation have to appear in
such a model as a zero mode equation, i.e. as an equation for the
massless Dirac field averaged over the complex Kerr string,
similar to the appearance of mass in the massless dual string
models.

\section{Other Aspects of the Extended Kerr Electron.}

{\bf Regularized source.} The twosheeted topology of the Kerr
geometry caused long discussion on the problem of Kerr source.
Series of the  papers has been devoted to an alternative approach
avoiding the twosheetedness.  Israel \cite{Isr} suggested to
truncate negative sheet along the disk $r=0,$ which resulted in a
disk-like source of the Compton size with a singular distribution
of matter $\delta(r).$ Subsequent investigations showed that such
a disk has to be rigidly rotating and built of  exotic
superconducting matter. An important correction was given by
L\'opez \cite{Lop}, who shifted the surface of the disk  to
$r=r_e=e^2/2m .$ The resulting source is the relativistically
rotating oblate ellipsoidal shell having Compton radius and the
thickness corresponding to classical radius of the electron $r_e$.
The resulting source turns out to be regularized; however it
contains a singular matter distribution on the shell. L\'opez
showed that gravity gives a very essential contribution to the
mass. The subsequent steps were related to the treatment of the
source in the form of a relativistically rotating bag filled by a
false vacuum \cite{BurBag,BEHM}. In such a model the local
gravitational field will be extremely small in all the points of
space-time which turns out to be really Minkowskian everywhere.
However, it was shown in \cite{BurReg} that, in spite of the very
small local contribution, {\it gravity possesses exclusively
strong  non-local effect}; in fact it provides regularization,
determining the point of phase transition from external (true)
electro-vacuum of the Kerr-Newman solution to the regular false
vacuum inside the bag. It is expected that such smooth
superconducting sources may be formed by Higgs fields in a
supersymmetric version of the $U(1)\times \tilde U(1)$ field
model. An image of corresponding regularized Kerr source is shown
on the fig.3.

\begin{figure}[ht]
\centerline{\epsfig{figure=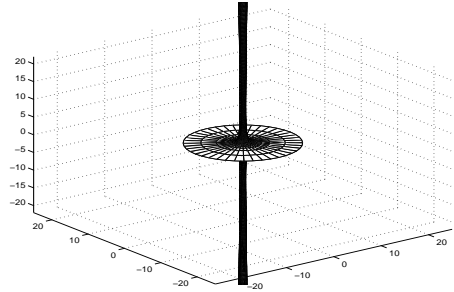,height=4cm,width=6cm}}
\caption{Kerr's electron dressed by Higgs field: relativistic disk
and two axial half-strings, carriers of the wave function.}
\end{figure}

{\bf Twistors and scattering.} One of the most problematic and
frequently asked questions concerns the seeming contradiction
between the large Compton size of  the Kerr electron and the
widespread statement on the point-like structure of electron
obtained in the experiments on deep inelastic scattering. The
explanation suggested in \cite{BurTwi}  is as follows. The
momentum of a massive particle is represented as a sum of the
lightlike parts $p_L^{\m}$ and $p_R^{\m}.$ For relativistic boosts
we have usually either $p_L << p_R$ or $p_L>>p_R$, which
determines the sign of helicity. As a result one of the axial
semi-strings turns out to be strongly dominant and another one
represents only a small correction, which allows to use the
perturbative twistor-string model for the scattering
\cite{Nai,Wit}, which is based on  a reduced description in terms
of the lightlike momentums and helicities. So, the relativistic
scattering is determined only by one of the axial half-strings.
One can conjecture that {\it the Kerr disk of the Compton size may
be observed only for polarized electrons in the low-energy
experiments with a very soft resonance scattering.}

One more question is related with the usefulness and/or necessity
of the twistor approach. First of all twistors are absolutely
necessary to determine exact form of the Kerr geometry in general
case apart from the case of a standard Kerr-Newman form. Second, a
very important application is related to discussed above
twistor-string theory for scattering at high energies.  Next,
there is evidence
 that twistors may play principal role for the space-time description of
 interactions at any energies. It follows from the treatment of the
  exact multiparticle
Kerr-Schild solutions which were obtained recently, using
generating functions of the Kerr theorem $F(Y),$ having higher
degrees in $Y .$ Forming the function $F$ as a product of a few
one-particle functions in the form of the known blocks $F_i(Y)$,
i.e. $ F(Y)\equiv \ \prod _{i=1}^k F_i (Y) ,$  one obtains
multiparticle solutions in which interaction between particles
occurs via a common singular twistor line \cite{Multiks}.

Finally, it should be noted that the associated with Kerr geometry
Minkowski space-time has indeed a twofold topology, and the usual
Fourier transform does not work for the functions formed by
complex shift. Thereby, the traditional for QED transform to the
momentum space cannot be performed in this case. Meanwhile, at
least the wave functions and S-matrix turns out to be well defined
by transform to twistors space \cite{Wit} which takes in some
sense an intermediate position between the coordinate an momentum
space. The corresponding twistor transform is obtained from
coordinate representation by a Radon transform \cite{GGV} which
represents a generalization of the usual Fourier transform.

The formalism based on twistor transform seems to be closely
adapted to the local and topological quantum field theories. In
particular, as it was shown in \cite{Wit}, the typical scattering
of plane waves is replaced with the scattering of the wave
functions having the support on singular twistor lines, which
seems to be very perspective for description of the strongly
polarized wave functions on the Kerr background.

\subsection{Acknowledgments}
Author wishing to acknowledge Organizing committee of the QFEXT'07
conference for very kind invitation and financial support. This
work was supported by RFBR grant 08-00234.

\end{document}